**Impact-driven planetary desiccation: the origin of the dry Venus**

Kosuke Kurosawa*

*Planetary Exploration Research Center, Chiba Institute of Technology, 2-17-1, Tsudanuma, Narashino, Chiba 275-0016, Japan*

*Corresponding Author

Kosuke Kurosawa

*E-mail address*: kosuke.kurosawa@perc.it-chiba.ac.jp

Tel: +81-47-478-0320

Fax: +81-47-478-0372

Street address: 421, 4th building, 2-17-1, Tsudanuma, Narashino, Chiba 275-0016, Japan

**Abstract**

The fate of surface water on Venus is one of the most important outstanding problems in comparative planetology. Although Venus should have had a large amount of surface water (like the Earth) during its formation, the current water content on the Venusian surface is only 1 part in 100000 of that of the mass of Earth's oceans. Here a new concept is proposed to explain water removal on a steam-covered proto Venus, referred to as "impact-driven planetary desiccation". Since a steam atmosphere is photochemically unstable, water vapor dissociates into hydrogen and oxygen. Then, hydrogen escapes easily into space through hydrodynamic escape driven by strong extreme ultraviolet radiation from the young Sun. The focus is on the intense impact bombardment during the terminal stage of planetary accretion as generators of a significant amount of reducing agent. The fine-grained ejecta remove the residual oxygen, the counter part of escaped hydrogen, via the oxidation of iron-bearing rocks in a hot atmosphere. Thus, hypervelocity impacts cause net desiccation of the planetary surface. I constructed a stochastic cratering model using a Monte Carlo approach to investigate the cumulative mass of nonoxidized, ejected rocks due to the intense impact bombardment. The ejecta mass after each impact was calculated using the π-group scaling laws and a modified Maxwell's Z model. The effect of projectile penetration into the ground on the ejecta mass was also included. Next, an upper limit on the total amount of removed water was calculated using the stoichiometric limit of the oxidation of basaltic rocks, taking into account the effect of fast $H_2$ escape. It is shown that a thick steam atmosphere with a mass equivalent to that of the terrestrial oceans would be removed. The cumulative mass of rocky ejecta released into the atmosphere reaches 1 wt% of the host planet, which is 10000 times of the current mass of the Earth's atmosphere. These results strongly suggest that chemical reactions between such large amounts of ejecta and planetary atmospheres are among the key factors required to understand atmospheric mass and its composition, not only in the Solar System but also in extrasolar systems. **(350 words)**

*Highlights:*

1. Impact-driven planetary desiccation is proposed to explain the water deficit on the Venusian surface.

2. Impact bombardment excavates the planetary crust/mantle down to several tens km.

3. Impact-generated ejecta remove EUV-generated oxygen in the hot Venusian atmosphere via high-temperature oxidation.

4. The surface water on Venus with a mass of the terrestrial ocean would be removed.

5. Impacts cause an efficient mixing between the crust/mantle and the atmosphere.

**1. Introduction**

The reason for the lack of moisture on Venus [e.g., *Donahue et al*., 1982] has been a long-standing problem in the context of understanding how habitable planets are formed. Since Venus is Earth's twin, it would have been formed with a similar amount of surface water to that of Earth's ocean [e.g., *Matsui and Abe*, 1986]. The total amount of surface water on planets is one of the most important parameters required to determine the evolution of the surface environment on terrestrial planets. This is because water largely controls geodynamical and geochemical processes owing to its strong greenhouse effect in the gas phase and its nature as a strong solvent in the liquid phase. In addition, water largely affects the rheological properties of rocks [e.g., *Korenaga and Karato*, 2008]. The mechanism responsible for causing a water deficit on the Venusian surface must be explored to understand the origin of the significant difference in the surface environment between Earth and Venus.

Previous studies proposed that the problem of water removal from Venus comes down to the fate of oxygen generated by extreme ultra violet (EUV) because of the following reasons. Water vapor might be dissociated into hydrogen and oxygen by the much stronger EUV radiation from the young Sun [e.g., *Kasting and Pollack*, 1983] compared with that at the present time. Hydrogen can easily escape into space through EUV-driven hydrodynamic escape [e.g., *Kasting and Pollack*, 1983; *Zahnle and Kasting*, 1986; *Sasaki*, 2008]. The Pioneer probe showed that the ratio of deuterium to hydrogen (D/H) in the current Venusian atmosphere is enriched by a factor of 120±40 [*de Bergh et al.*, 1991]. Such high D/H ratio is thought to be an evidence of the hydrogen escape to the space [e.g., *Kasting and Pollack*, 1983]. In the context of the dissociation of water vapor into hydrogen and oxygen, previous studies have investigated the fate of oxygen based three processes: (1) frictional escape caused by drag-off by the escaping hydrogen flow [e.g., *Chassefière*, 1996; *Gillmann et al.*, 2009; *Kasting and Pollack*, 1983; *Sasaki*, 2008; *Zahnle and Kasting*, 1986], (2) nonthermal escape by ion pick-up owing to the solar wind [e.g., *Kulikov et al.*, 2007; *Terada et al.*, 2009] and (3) oxidation of the Venusian surface [e.g., *Hamano et al.*, 2013; *Gillmann et al.*, 2009; *Lewis and Prinn*, 1984].

The first and second processes are considered equivalent to the escape of oxygen into space. A serious problem as regards the first process is the accumulation of oxygen that is left behind the escaping hydrogen flow [*Sasaki*, 2008]. According to recent numerical calculations of multi-component hydrodynamic flows [*Sasaki*, 2008], the escape flux of hydrogen is higher than that of oxygen. Thus, EUV-generated oxygen is inevitably accumulated in the atmosphere, eventually leading to a shutdown of the hydrodynamic escape itself [*Sasaki*, 2008]. The escape amount of oxygen by means of the second process depends strongly on the activity of the ancient Sun, which is highly uncertain [e.g., *Kulikov et al.*, 2007]. In addition, more accurate evaluation of interactions between charged particles in the solar wind and planetary atmospheres, using a three-dimensional multi-species magnetohydrodynamic (MHD) model, indicates that the escape rate of oxygen is much lower than previously thought [*Terada et al.*, 2009].

A serious problem associated with the third process is the lack of a driving force for surface replacement [e.g., *Gillmann et al.*, 2009]. The upper limit of oxygen removal caused by surface oxidation is estimated to be 0.1 times than an amount of surface water equivalent to that of the terrestrial oceans (hereafter referred to as TO; $1.4 \times 10^{21}$ kg), based on the high average rate of volcanism over Venus' history [e.g., *Gillmann et al.*, 2009]. Recently, *Hamano et al.* (2013) proposed a new idea, namely that atmospheric oxygen could be removed through oxidation of a giant-impact-induced magma ocean. They showed that terrestrial planets are divided into two types, Earth-like ocean-covered planets (Type-I) and Venus-like dry planets (Type-II), a difference which appears to be correlated with distance from the planet's host star. If Venus belongs to the latter, "Type-II" planets, such a magma ocean (if affected by active thermal convection) may behave as a massive sink for atmospheric oxygen. However, the semi-major axis of the current orbit of Venus, 0.72 au, is very close to that which defines the boundary between the two types of planet, 0.6–0.8 au. Thus, it is not clear whether or not Venus belongs to the Type II planetary category [*Hamano et al.*, 2013].

Here I propose a new mechanism, called "impact-driven planetary desiccation", to explain the water loss from steam-covered proto-Venus. A clue to the fate of surface water on Venus lies in those many impact craters on the Moon and Mars. Venus should have suffered similar impact bombardment at the same time as those on the Moon and Mars. I focus on such intense impact bombardment during the terminal stage of planetary accretion [e.g., *Bottke et al.*, 2010] as the generator of reducing agents on planetary surfaces. Hypervelocity impacts onto proto-Venus during the terminal stage of planetary accretion are expected to have released a significant amount of fine-grained, highly reactive rocky materials into the atmosphere. At the same period, the accumulation of oxygen would proceed on proto-Venus via the photodissociation of the steam atmosphere and subsequent hydrogen escape into space as discussed above. The ejected rocks due to hypervelocity impacts are expected to remove the accumulated oxygen via the oxidation of iron-bearing rocks, i.e., $2FeO(s) + 1/2O_2(g) \rightarrow Fe_2O_3(s)$. As a result, hypervelocity impacts cause net desiccation of the atmosphere via oxidation of the planetary crust/mantle.

The objective of this study is to assess the significance of impact-driven planetary desiccation on steam-covered proto-Venus quantitatively based on impact physics and the impactor population at the terminal stage of planetary accretion. I calculated the cumulative mass of rocky ejecta on proto-Venus based on a number of previous studies on impact-driven excavation. Then, the possible amount of removed water is discussed.

## 2. Numerical model

I constructed a numerical model using a Monte Carlo approach to include the effects of the stochastic nature of impact bombardment from the size, velocity and angle distributions of impactors during the terminal stage of planetary accretion. The excavated mass and depth after each impact event were calculated analytically. Then, the cumulative mass of non-oxidized ejecta through the intense impact bombardment was calculated. In this section, I describe the details of the model. The capability of rocky ejecta for oxygen removal is discussed in Section 2.1. The procedure to calculate the excavated mass at each impact event and the size of ejected particles are presented in Section 2.2. The reaction cross-section of rocky ejecta is an important factor to discuss the oxygen removal efficiency of rocky ejecta. The size and velocity distribution of impactors at the terminal stage of planetary accretion is described in Section 2.3. I discuss how to calculate the possible amount of removed oxygen at the end of the impact bombardment in Section 2.4. The effect of nonrenewable oxidized deposits is included in the calculation.

**2.1 The oxygen removal efficiency**

The water-removal efficiency ($\alpha$) of impact-generated ejecta, including the effect of hydrogen escape, was calculated stoichiometrically at 10.6 g kg$^{-1}$ based on the average FeO content on the Venusian surface [*Surkov et al.*, 1984, 1986], 8.6 wt%. To remove an amount of surface water equivalent to that of the terrestrial oceans (TO; 1.4 × 10$^{21}$ kg), an impact-generated ejecta mass of 1.3 × 10$^{23}$ kg, corresponding to a basaltic crust/mantle of 90 km, is required.

For the calculations, it is assumed that ferrous iron is oxidized to ferric iron via the following oxidation reactions,

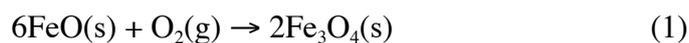

$$6FeO(s) + O_2(g) \rightarrow 2Fe_3O_4(s) \qquad (1)$$

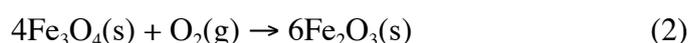

$$4Fe_3O_4(s) + O_2(g) \rightarrow 6Fe_2O_3(s) \qquad (2)$$

The net reaction is described by $4FeO(s) + O_2(g) \rightarrow 2Fe_2O_3(s)$. Thus, it is necessary to address whether the above reaction proceeds in the forward direction. The chemical composition of the Fe–H–O system in thermal and chemical equilibrium was investigated using the Lewis code [*Gordon and McBride*, 1994]. Figure 1 shows the oxidation state of iron in the presence of excess oxygen as a function of the molar mixing ratio of oxygen to water vapor and temperature. As mentioned in Section 1, photodissociation of water vapor and subsequent $H_2$ escape would lead to a gradual increase in the partial pressure of molecular oxygen in an early Venusian atmosphere [e.g., *Gillmann et al.*, 2009]. Thus, the molar mixing ratio of molecular oxygen to water vapor was varied from 0 to 0.7 in the calculations. The total pressure in the system was set to 300 atm. This value is expected for the surface pressure in a steam atmosphere with a mass of 1 TO in the presence of Venus' gravity. Ferric iron ($Fe_2O_3$) is stable if the molar mixing ratio of $O_2$ to $H_2O$ exceeds a value of 0.5 for a wide range of temperatures. Although results for a single value of the pressure (300 atm) are shown here, the results do not strongly depend on pressure. The high-temperature boundary of the stability field for $Fe_2O_3$ decreases slightly, from 1900 K to 1600 K, if 1 atm is used for the total pressure. If the molar mixing ratio of $O_2$ to $H_2O$ does not exceed 0.5, only the oxidation reaction described by Eq. (1) occurs, resulting in a reduction of the water-removal efficiency, $\alpha$, from 10.6 g kg$^{-1}$ to 7.1 g kg$^{-1}$. Nevertheless, remote-sensing data [*Pieters et al.*, 1986] supports the presence of ferric minerals on the Venusian surface. In this section, I only demonstrate that the oxidation reactions of Eqs (1) and (2) proceed in the forward direction in the presence of EUV-generated excess oxygen. Additional discussion, including that of the kinetics of the oxidation reactions, is presented in Section 4.1.

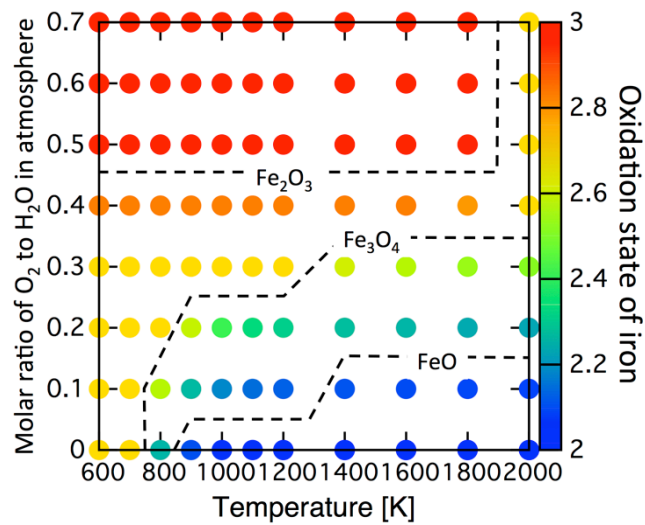

**Figure 1.** Oxidation state of iron as a function of the molar ratio of $O_2$ to $H_2O$ in a steam atmosphere and of temperature. The oxidation states of FeO, $Fe_3O_4$, and $Fe_2O_3$ correspond to 2, 8/3, and 3, respectively. The elemental ratio of H to Fe was fixed at H:Fe = 2:1. The total pressure was set to 300 atm.

## 2-2. Impact calculations

The excavated rocky mass and depth are calculated analytically using a combination of the π-group scaling laws [e.g., *Schumidt and Housen*, 1987] and a modified Maxwell's Z-model [e.g., *Croft*, 1980]. The details of the calculations are presented in Appendix A. Figures 2a and 2b shows typical examples of the geometry of a transient crater and of the excavated region for impactors with the size (a) 5 km and (b) 50 km in diameter. The ejected mass and the excavation depth, both as a function of impactor diameter, are shown in Figure 3a. The ejected mass reaches 10–100 times that of the impactor mass and the excavation depth is comparable to the projectile diameter.

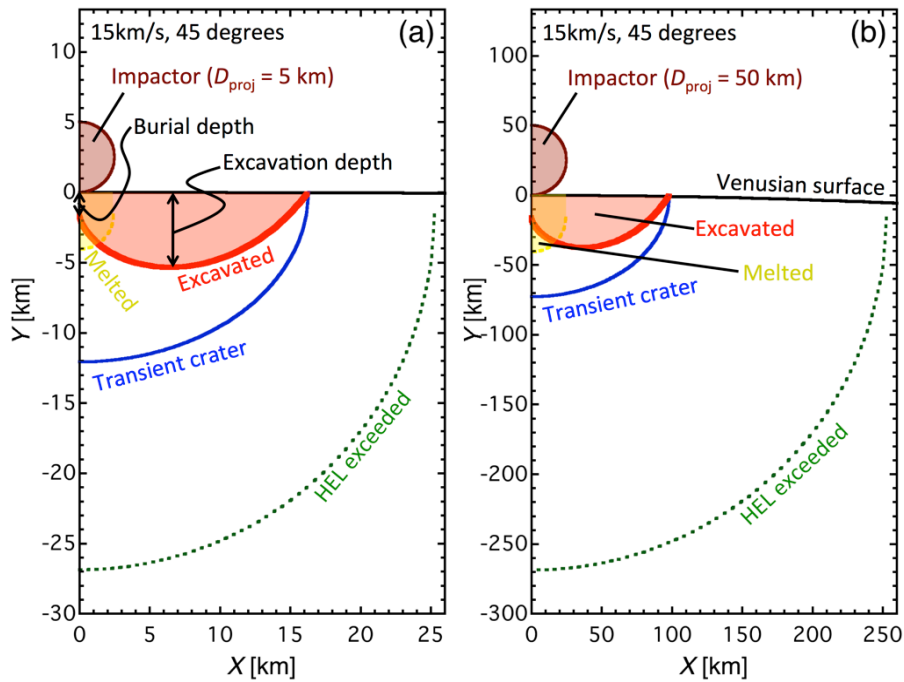

**Figure 2.** Typical examples of the geometry of crater formation for a projectile diameter of (a) 5 km and (b) 50 km, impacting at 15 km s$^{-1}$ under an angle of 45°. The excavated region is shown as the hatched area.

It is assumed that impact ejecta react efficiently with an atmosphere up to the stoichiometric limit of the oxidation reaction, $2FeO(s) + 1/2 O_2(g) \rightarrow Fe_2O_3(s)$, as discussed in Section 2.1. Thus, highly comminuted or melted ejecta are necessary to efficiently remove the accumulated oxygen. I calculated the size of the region compressed by more than the Hugoniot elastic limit (HEL; $P_{HEL} \sim 5$ GPa [e.g., *Melosh*, 1989]) and the pressure required for incipient melting ($P_{IM} \sim 100$ GPa [e.g., *Melosh*, 1989]) as follows. The pressure field, $P(r)$, generated by a hypervelocity impact as a function of the radial distance from the impact point, $r$, is obtained from the size of the isobaric core [*Pierazzo et al.*, 1997], $R_{ic}$, and the pressure-decay exponent [*Pierazzo et al.*, 1997], $n$, as [e.g., *Melosh*, 1989]

$$P(r) = P_o \left(\frac{r}{R_{ic}}\right)^{-n} \text{ for } (r > R_{ic}), \tag{3}$$

where

$$P_o = \rho_o V_s u_p, \tag{4}$$

$$V_s = C_o + s u_p, \tag{5}$$

$$u_p = \frac{v_{impact} \sin(\vartheta_{impact})}{2}, \tag{6}$$

$$\log_{10}(R_{ic}) = -0.428 + 0.0256 \times \log_{10}[v_{impact} \sin(\theta_{impact})] \text{ and} \tag{7}$$

$$n = -0.31 + 1.15 \times \log_{10}[v_{impact} \sin(\theta_{impact})], \tag{8}$$

where $P_o$, $\rho_o$, $V_s$, $u_p$, $C_o$ and $s$ are the shock pressure in the isobaric core, the reference density, shock velocity, particle velocity, the bulk sound velocity and a constant, respectively. In this formulation, the one-dimensional impedance-matching solution [e.g., *Melosh*, 1989] is used to obtain the shock pressure in the isobaric core as a first-order estimate and assumed collisions of identical materials. Since I considered oxidation of basaltic rocks, I used parameters pertaining to basalt [*Melosh*, 1989], $\rho_o$ = 2860 kg m$^{-3}$, $C_o$ = 2600 m s$^{-1}$ and $s$ = 1.62. Eqs (7) and (8) were obtained by hydrocode calculations as the same as Eq. (A2). Consequently, the radii of the HEL-exceeded and melted regions, $R_{HEL}$ and $R_{melt}$, are, respectively,

$$R_{HEL} = R_{ic} \left(\frac{P_o}{P_{HEL}}\right)^{\frac{1}{n}} \text{ and} \qquad (9)$$

$$R_{melt} = R_{ic} \left(\frac{P_o}{P_{melt}}\right)^{\frac{1}{n}}. \qquad (10)$$

Figure 3b shows a comparison of the impact-related lengths at 15 km s$^{-1}$ and for an impact angle of 45 degrees, including the radius of the transient crater and the radii of the HEL-exceeded and melted regions, calculated based on Eqs (A1), (9) and (10), respectively. This comparison shows that the full body of ejecta suffers from shock pressures in excess of $P_{HEL}$ and that the ejecta are expected to be highly comminuted.

    The sizes of the HEL-exceeded rocky ejecta are expected to be similar to

those of lunar regolith particles, because they are also produced by frequent hypervelocity impacts onto the Moon [e.g., *Melosh*, 2011]. The typical grain size of regolith particles is 45–100 μm [e.g., *McKay et al.*, 1991]. Since such fine-grained ejecta have a significant reaction cross-section, the water-removal efficiency may reach the stoichiometric limit. Further discussion of the kinetics is presented in Section 4.1, based on results of combustion experiments using a $FeO–Fe_3O_4$ powder with a similar size to that of lunar regolith. The radius of the melted region is much smaller than that of the transient crater, except for large impactors (which are rare in the present calculations). Thus, the contribution from impact melt to water removal was neglected in this study.

**2-4. Impactor size and velocity distributions**

It was assumed that dynamic transport of impactors occurred from the inner asteroid belt to the early Venus. Since the impactor population onto Venus during the terminal stage of planetary accretion has not been investigated, this study employs the size [*Bottke et al.*, 2005, 2010] and velocity distributions [*Ito and Malhotra*, 2006] of impactors on Earth. The size–frequency distribution (SFD) is given by $N(>D) = CD^{-q}$, where $N$, $D$, $C$, and $q$ are the cumulative number of impactors larger than a given diameter, the diameter of the impactor, a constant, and an exponent, respectively. *Bottke et al.* (2005) investigated the evolution of the SFD by including the effects of collisional fragmentation of main-belt asteroids and found that bodies originally larger than 120 km are expected to stave off fragmentation due to mutual collisions in the main asteroid belt. The estimated size exponent $q$ for bodies > 120 km is 1 [*Bottke et al.*, 2010]. This SFD for large bodies is consistent with the SFD of current main-belt asteroids and the estimated projectile SFD for Martian impact basins [*Bottke et al.*, 2010]. The value of the size exponent $q$ for bodies < 120 km at the terminal stage of planetary accretion is still controversial. If the SFD of main-belt asteroids at that time was in a steady state owing to collisional cascade, $q$ should be 2.5 [*Tanaka et al.*, 1996]. Thus, I employed $q$ = 2.5 as a typical value and varied $q$ from 2.1 to 2.8.

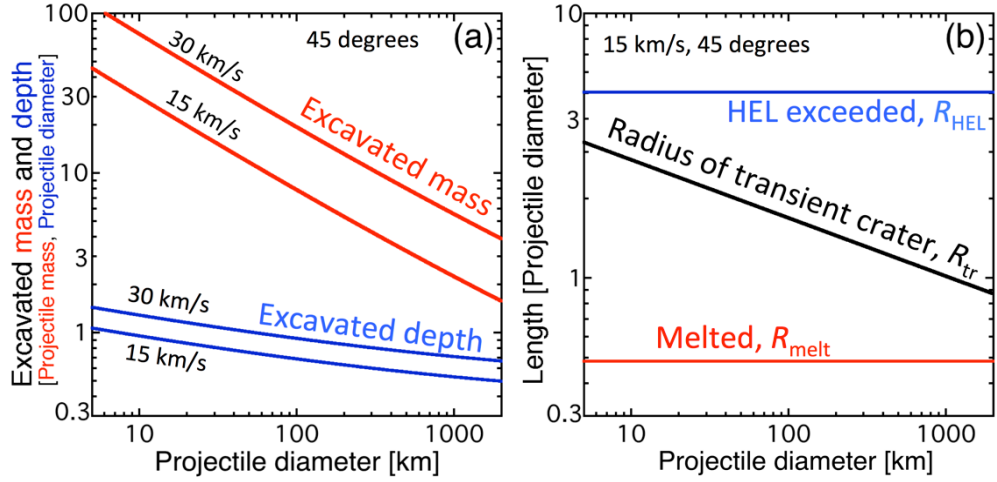

**Figure 3.** (a) Excavated mass and depth following each impact, normalized by projectile mass and diameter, as a function of projectile diameter and impact velocity. (b) Comparison of impact-related lengths, including the radius of the HEL-exceeded hemisphere, $R_{HEL}$ (Eq. 9), the radius of the melted hemisphere, $R_{melt}$ (Eq. 10), and the radius of transient craters, $R_{tr}$ (Eq. A1), as a function of projectile diameter. All lengths are normalized by projectile diameter.

The distribution of impact velocities from the asteroid belt to Earth is obtained from an *N*-body orbital calculation under the influence of the gravitational field of the Sun and nine planets [*Ito and Malhotra*, 2006]. Figures 4a and b show the size and velocity distributions of impactors used in this study. In general, the impact velocity onto Venus is higher than that onto Earth because of the difference in gravitational potential energy from the Sun between the two planets. Thus, the calculations provide a conservative estimate of the mass of fine-grained ejecta from Venus.

The distribution of impact angle, $\theta_{impact}$, is given by $\sin(2\theta_{impact})$ [*Shoemaker*, 1963]. The total impactor mass $M_{total}$ during the terminal stage of planetary accretion onto Earth has been estimated [e.g., *Chyba*, 1991; *Dauphas and Marty*, 2002; *O'Neill and Palme*, 1998] to range from $0.7 \times 10^{22}$ kg to $4.2 \times 10^{22}$ kg. Since Venus may have experienced a similar impact bombardment at the same time, the total impactor mass was varied from $1 \times 10^{22}$ kg to $5 \times 10^{22}$ kg. The effect of stochastic bombardment of large impactors was taken into account by employing a Monte Carlo approach. The Mersenne Twister algorithm [*Matsumoto and Nishimura*, 1998] was used in the Monte Carlo calculations. The maximum diameter of impactors, $D_{max}$, is varied from 500 km to 2000 km. This range of $D_{max}$ was chosen because such $D_{max}$ values were estimated in previous studies pertaining to catastrophic impact events on Earth [e.g., *Sleep and Zahnle*, 1998; *Bottke et al.*, 2010]. The cut-off diameter for small bodies, $D_{min}$, was set to 5 km. This value is much smaller than the crust thickness required removing an amount of surface water equivalent to 1 TO as discussed in Section 2.1.

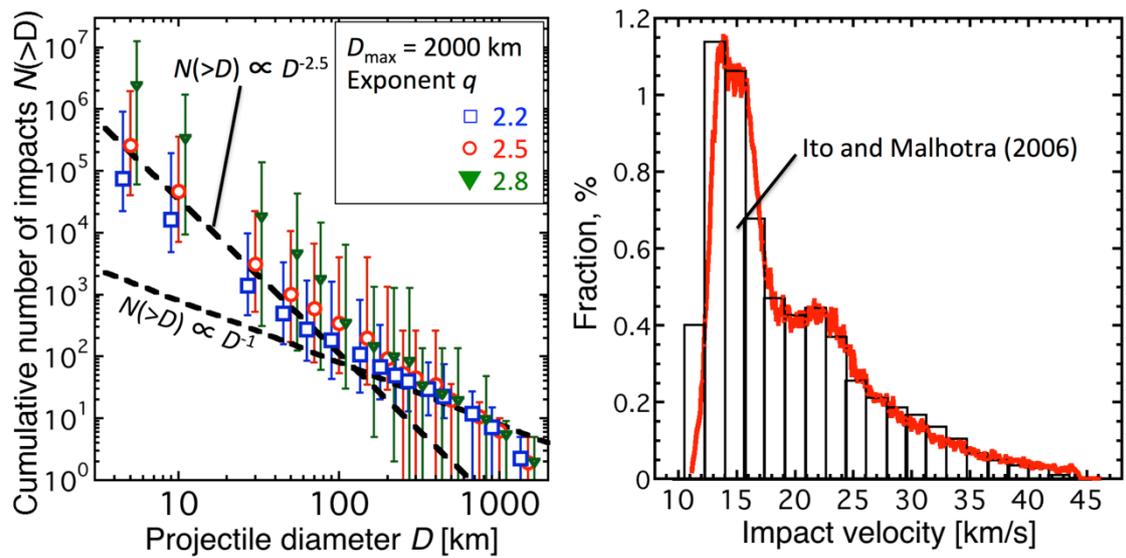

**Figure 4.** (a) Size distribution of impactors during the terminal stage of planetary accretion used in this study. Cumulative numbers of impacts for a given diameter are plotted. Three values of the exponent $q$ for impactors < 120 km were used. The maximum diameter, $D_{max}$, was set to 2000 km. The black dotted lines are the scaled input distributions for large (>120 km) and small (<120 km) impactors. The data scatter originates from the stochastic nature of the bombardment. The error bars are obtained from the 1σ dispersion of the results of the Monte Carlo runs (see main text). (b) Velocity distribution of impactors. The histogram shows the scaled input distribution of impact velocity.

**2-5. Calculation of the cumulative mass of nonoxidized fresh ejecta**

Treatment of nonrenewable oxidized deposits on the surface is key to quantitatively assess the cumulative amount of water vapor that was removed. Since oxidized deposits cannot contribute to oxygen removal during subsequent impact events, an estimate of the cumulative mass of nonoxidized ejecta, $M_{fresh}$, is required to investigate the significance of impact-driven planetary desiccation. The post-impact behavior of fine-grained ejecta in a thick atmosphere, however, is significantly complicate and has not been fully understood [e.g., *Schultz*, 1992a, b]. Consequently, it is difficult to calculate the time evolution of the degree of oxidation of the Venusian crust/mantle because the change in the degree of mixing between oxidized ejecta and nonoxidized "fresh" rocks during impact bombardment is unknown.

In this study, $M_{fresh}$ was approximately estimated based on the concept of "impact-induced convection" [*Senshu et al.*, 2002], as follows. Hypervelocity impacts cause excavation and thus break the isostatic equilibrium of the crater cavity. An amount of deep "fresh" rocks comparable to that of the excavated "stale" rocks below the excavated region is expected make its way to the planetary surface through isostatic recovery. This "impact-induced convection" will lead to replenishment of the planetary surface. Given that the planetary surface is well mixed because of the occurrence of large numbers of impact-driven excavation events and subsequent uplift, a detailed treatment of the redistribution of oxidized impact ejecta in a thick atmosphere and their history is not needed in order to assess the cumulative amount of oxidized rock at the end of the impact bombardment. Thus, the evolution of the excavated fraction was recorded as a function of depth from the Venusian surface after each impact during the calculation, rather than the history of ejecta; i.e., rather than as a function of ejecta movement and mixing between stale and fresh rocks.

The cumulative mass of nonoxidized, "fresh" ejecta, $M_{fresh}$, was calculated using the excavated fraction as a function of depth from the Venusian surface, $h$. For simplicity, it was assumed that the shape of the excavated cavity is part of a spherical shell with thickness $H_{exc}$ and covering a volume $V_{exc}$, calculated based on Eqs (A5) and (A6), respectively. Then, the solid angle of the transient crater, $d\Omega$, from the centre of Venus was calculated as

$$d\Omega = \frac{4\pi V_{exc}}{\frac{4}{3}\pi \left[R_{Venus}^3 - (R_{Venus}-H_{exc})^3\right]} \tag{11}$$

where $R_{Venus}$ = 6052 km is the radius of Venus. The surface area of the excavated cavity at a given depth $S(h)$ is expressed as

$$S(h) = \left(R_{\text{Venus}}^2 - h^2\right) d\Omega \quad (h < H_{\text{exc}}). \tag{12}$$

Thus, the excavated fraction as a function of depth from the Venusian surface, $f(h)$, after the $i^{\text{th}}$ impact is given by

$$f(h)_i = f(h)_{i-1} + \frac{d\Omega_i}{4\pi} \quad (h < H_{\text{exc}, i}) \tag{13}$$

and

$$f(h)_i = f(h)_{i-1} \quad (h > H_{\text{exc}, i}), \tag{14}$$

where $H_{\text{exc},i}$ is the excavation depth of the $i^{\text{th}}$ impact. Note that overlaps of different craters were not considered in this study. If $f$ exceeds unity, $f$ is fixed to 1 following the calculation. Finally, the cumulative mass of nonoxidized fresh ejecta was calculated as

$$M_{\text{fresh}} = \int_0^{R_{\text{Venus}}} 4\pi \rho_a f(h)(R_{\text{Venus}}^2 - h^2)\, dh. \tag{15}$$

For example, if $f$ becomes unity down to 10 km, it is approximated by the formation of the highly-oxidized deposit with the thickness of 10 km.

## 3. Calculation results

I conducted 60 runs adopting the same values of $M_{total}$, $D_{max}$, and $q$. The calculated results, including $M_{fresh}$, the total number of impacts, $N_{total}$, and the total coverage of the transient craters on Venus, $\Sigma S_{tr}$, exhibit large scatter due to the stochastic nature of the bombardment. I extracted the data within the 1σ dispersion of the results of the 60 runs. Typically, half of the data points were rejected during the data reduction. Next, I calculated the average value of the extracted data. The 1σ dispersion of each quantity was used as the error bar. Figures 4a and 4b shows the size and velocity distribution of impactors in the calculation. In the calculations shown in Fig. 4a, $M_{total}$ of $4 \times 10^{22}$ kg and $D_{max}$ of 2000 km were used. The total number of impacts $N_{total}$ becomes $10^6$ in the calculation although it depends on $D_{max}$ and $q$ and has a large scatter.

Figure 5 shows the excavated fraction $f(h)$ at the end of the bombardment as a function of depth from the surface. The Venusian surface is completely excavated during the impact bombardment, to 10–80 km below the surface when $M_{total}$, $q$, and $D_{max}$ are set to $4 \times 10^{22}$ kg, 2.5, and 1000 km, respectively. In addition, the deep interior down to 100–1000 km below the surface is partially excavated by a small number of large impactors. Such large impact events are expected to transport nonoxidized rocks from the deep interior to the Venusian surface. In contrast, a significant number of small impacts would contribute to mixing of the surface rocks and the hot Venusian atmosphere.

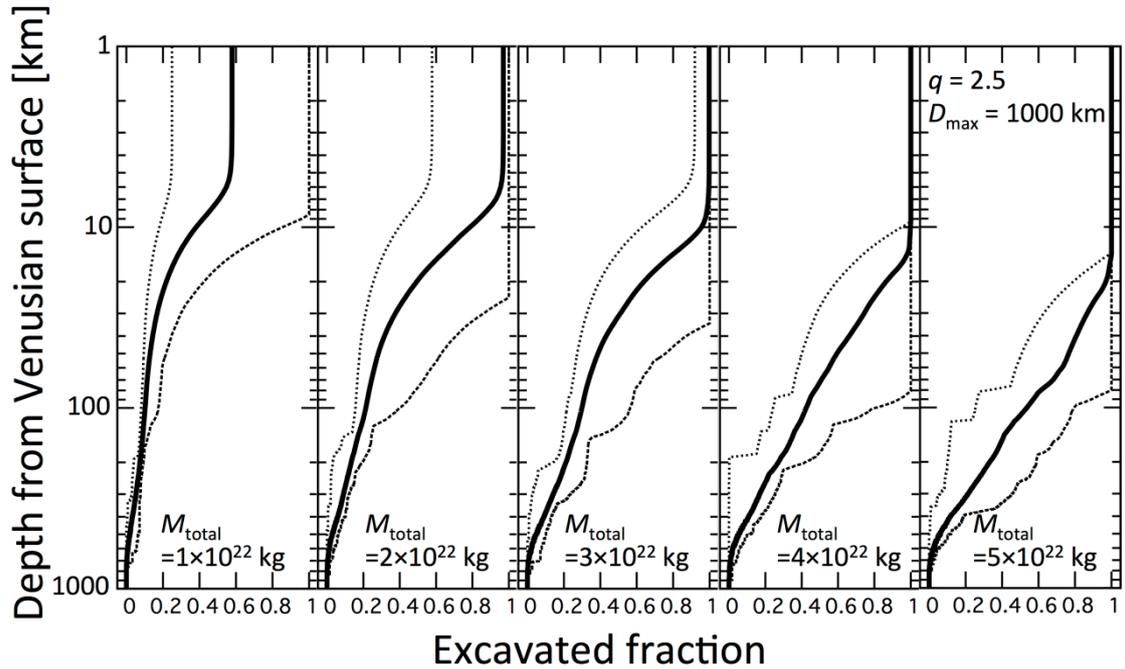

**Figure 5.** Excavated fraction at the end of impact bombardment as a function of depth from the surface. The total impactor mass used is shown. Results for $q = 2.5$ and $D_{max} = 1000$ kg are shown. The bold curves are the average values and the dotted curves are 1$\sigma$ uncertainties (see main text).

Figure 6 shows a summary of the Monte Carlo runs pertaining to the amount of water removal, $M_{H_2O} = \alpha M_{fresh}$, as a function of $M_{total}$, $q$, and $D_{max}$. Although the results are affected by significant scatter owing to the stochastic nature of the impact bombardment, $M_{H_2O}$ increases with increasing $M_{total}$. If $M_{total}$ exceeds $2 \times 10^{22}$ kg, 0.5 wt% of the mass of Venus, the oxygen accumulated in >0.6 TO of water can be removed. The previous studies showed that the water vapor of 0.1-0.3 TO has been possibly removed by three processes as discussed in Section 1 [e.g., *Kulikov et al.*, 2007; *Gillmann et al.*, 2009; *Zahnle and Kasting*, 1986]. Thus, impact-driven planetary desiccation would be the main contributor of the water removal on the Venus. These results show that the impact bombardment during the terminal stage of planetary accretion is a missing term in the driving force of surface replacement on proto-Venus.

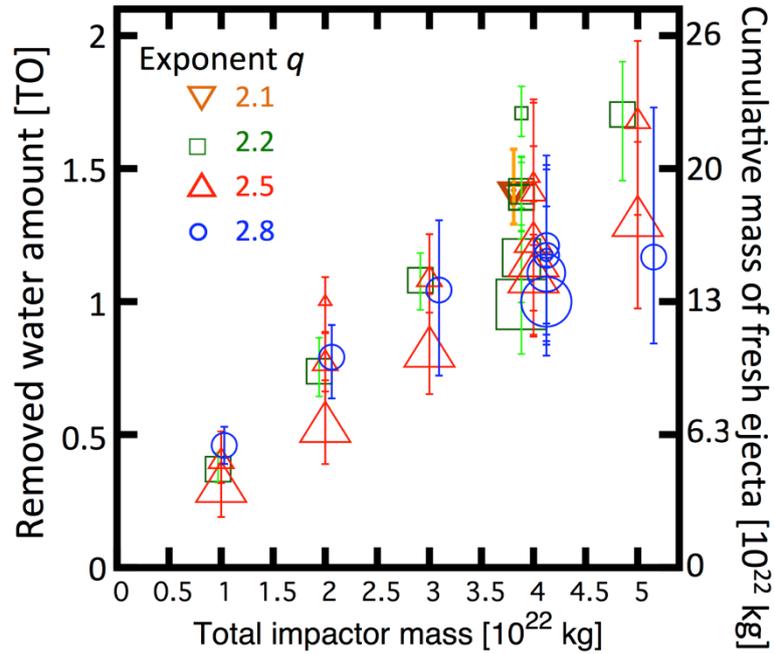

**Figure 6.** Possible total amount of removed water, $M_{H2O}$, as a function of total impactor mass, $M_{total}$, size exponent, $q$, and the maximum impactor diameter, $D_{max}$. Four different symbol sizes are used, corresponding to $D_{max}$ are 500 km, 1000 km, 1500 km, and 2000 km. For reasons of clarity, $M_{total}$ of each point is artificially shifted, depending on the exponent $q$. The cumulative mass of nonoxidized "fresh" ejecta is also shown on the right axis.

## 4. Discussion

### 4.1 Validity and limitations of the model

In this section, I discuss the oxidation sequence of fine-grained rocky ejecta in the thick and hot Venusian atmosphere. Based on the stability of ferric iron in the presence of molecular oxygen and under the prevailing impact physics, I have demonstrated that, during the terminal stage of planetary accretion, a sufficient amount of nonoxidized rocks is generated to consume the EUV-generated oxygen in water vapor with a mass of > 0.6 TO, if $M_{total}$ is larger than $2 \times 10^{22}$ kg (Section 3). Note that the above estimate is an upper limit on the total amount of oxygen that can be removed, because the water-removal efficiency $\alpha$ used in this study is estimated based on the stoichiometric limit of the oxidation reactions, Eqs (1) and (2). The oxidation kinetics of fine-grained rocky ejecta in the thick and hot Venusian atmosphere should be considered to obtain an accurate estimate of the amount of removed oxygen. Although a quantitative calculation of $\alpha$ based on the oxidation kinetics and the dynamics of impact-generated ejecta under the thick Venusian atmosphere has not yet been performed at this stage, I will next discuss the validity and limitations of the model.

*4.1.1. Timescale of high-temperature oxidation of iron-bearing minerals*

*Monazam et al.* (2014) conducted a series of high-temperature oxidation experiments of $FeO$–$Fe_3O_4$ powder in air (a mixture of $N_2$ and $O_2$ gas) using gravimetric measurements. The average size of their initial samples was 200 μm, which is larger than the average size of the HEL-exceeded rocks, 45–100 μm, as discussed in Section 2.2. The temperatures investigated in their experiments ranged from 1023 K to 1173 K, which can be used as an approximate value of the surface temperature of a steam-covered Venus [e.g., *Matsui and Abe*, 1986]. They reported that the conversion from FeO–$Fe_3O_4$ powder to $Fe_2O_3$ powder occurs within ~30 min at 1023–1173 K. The required time for the complete oxidation of FeO–$Fe_3O_4$ powder is likely much shorter than the residence time of the fine-grained ejecta in the thick Venusian atmosphere (see Section 4.1.2). They found that the rate-controlling process of the oxidation of the FeO–$Fe_3O_4$ mixture is oxygen transport through the hematite layer produced. The activation energy, $E_a$, of the rate-controlling step has also been obtained, 53.58 ± 3.56 kJ mol$^{-1}$. If I extrapolate $E_a$ at 1023–1173 K to the typical surface temperature of the present-day Earth, 300 K, the ratio of the reaction rate on Earth to that on steam-covered Venus reaches $10^{-6}$. Thus, oxidation of impact ejecta during their flight in the Earth's atmosphere would not occur if the extrapolation of $E_a$ to a lower temperature were valid. According to a geochemical analysis of the ejecta deposit layer on the present-day Earth, rocks collected a few 100 m below the surface are typically reduced more than surface rocks [*e.g., Navarre-Sitchler and Brantley*, 2007], suggesting that the impact ejecta were not oxidized during their flight through the Earth's atmosphere. Note that the activation energy of the chemical weathering of basalt through the interaction with the $N_2$–$O_2$ atmosphere under the conditions of the current Earth yields 70 ± 20 kJ mol$^{-1}$ for a range of temperatures, from 277 K to 323 K [*Navarre-Sitchler and Brantley*, 2007]. This value is similar to the $E_a$ of high-temperature oxidation of iron-bearing minerals, suggesting that the basalt oxidation rate on Venus may occur 10$^6$ times faster than that on the present-day Earth, if this extrapolation to a higher temperature is valid. Although I could not accurately estimate the oxidation rate of the fine-grained ejecta in the hot $H_2O$–$O_2$ atmosphere on early Venus, the required time for oxidation of the fine-grained ejecta in the early Venusian hot environment is likely much shorter than that on the present-day Earth.

*4.1.2. Ejecta retention time in a thick atmosphere*

The residence time of silicate particles with sizes smaller than 100 μm in the Venusian atmosphere can be approximately estimated using Stoke's law [e.g., *Melosh*, 2011]. The terminal velocity of silicate particles in the current Venusian atmosphere for 10–100 μm particles is listed in Table 9.1 of *Melosh* (2011). I derived an empirical equation for the terminal velocity, $v_{terminal}$, based on the listed values, i.e., $v_{terminal} = 4 \times 10^{-5} D_{particle}^{-2}$, where $D_{particle}$ is the diameter of a falling particle in micrometer. Figure 7 shows the residence time of silicate particles in the Venusian atmosphere as a function of $D_{particle}$ and the initial altitude of particle fall, $h_{fall}$. The residence time for $D_{particle}$ = 45–100 μm-size silicate particles exceeds 1 hour for a wide range of $D_{particle}$ and $h_{fall}$.

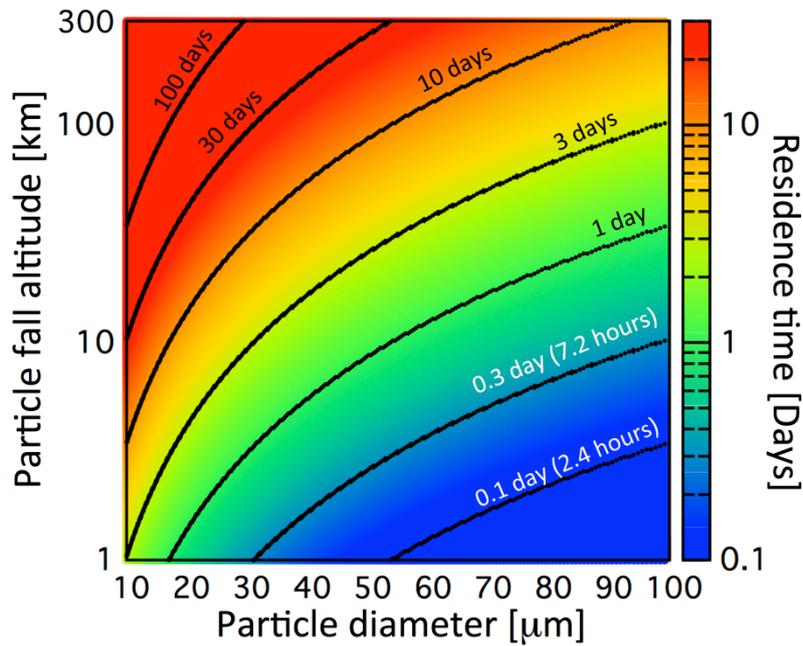

**Figure 7.** Residence time of silicate particles in the Venusian atmosphere as a function of particle diameter and the initial altitude of particle fall.

The above estimate is expected to be conservative for the residence time of the fine-grained ejecta in a thick atmosphere, because the residence time of silicate particles in the Venusian atmosphere after hypervelocity impacts might not be controlled by particle settling at the terminal velocity. It could be controlled by the hydrodynamic response of the atmosphere surrounding the crater after the hypervelocity impact [*Schultz*, 1992a, b]. *Schultz* (1992a, b) studied a time sequence of impact-generated winds in the Venusian atmosphere. The atmosphere above the impact point is blown away immediately by a rapidly expanding vapor plume. Next, a strong recovery wind flows into the crater and the fine-grained ejecta are entrained in the wind.

Thus, the actual residence time of the HEL-exceeded ejecta in the Venusian atmosphere may be much longer than that shown in Figure 7. Therefore, the fine-grained ejecta are expected to be oxidized during their retention in the hot atmosphere upon hypervelocity impact if the atmospheric mass that can interact with the ejecta during the event is sufficient for the FeO in the fine-grained ejecta to be completely oxidized to $Fe_2O_3$. The atmospheric mass available for oxidation is discussed in the next section

*4.1.3. Excavated mass versus atmospheric mass*

The ejection velocities of HEL-exceeded rocks are typically less than 10% of the impact velocity, because material ejection after impact is mainly driven by the "residual" particle velocity after adiabatic release [*Melosh*, 1989]. Thus, the HEL-exceeded ejecta would not disperse into the entire Venusian surface. Although a 2–4 mm-thick global ejecta layer was found at the K–Pg boundary on Earth [e.g., *Smit*, 1999], it mainly comes from an impact-generated vapor plume [e.g., *Artemieva and Morgan*, 2009], as opposed to the HEL-exceeded ejecta discussed in this study. The atmospheric mass above the tangent plane of the impact point, $M_{tangent}$, may be used as an upper limit to the atmospheric mass available for silicate oxidation. The available atmospheric mass $M_{tangent}$ is given by [e.g., *Melosh and Vickery*, 1989]

$$M_{tangent} = \frac{2\pi P_o}{g} H R_{Venus}; \tag{16}$$

$$H = \frac{RT}{\mu g}, \tag{17}$$

where $P_o$, $H$, $R$ (= 8.314 J mol$^{-1}$ K$^{-1}$), $T$, and $\mu$ are the surface pressure on Venus, the atmospheric scaleheight, a gas constant, the atmospheric temperature, and the molecular weight of the atmosphere, respectively. I assumed that the atmosphere is in hydrostatic equilibrium and isothermal. If $P_o$ = 300 atm, $T$ = 1000 K, and $\mu$ = 18 g mol$^{-1}$, $M_{tangent}$ = 7 × 10$^{18}$ kg. Oxidation of all ejecta is unlikely to occur if the excavated mass is much larger than $M_{tangent}$. In other words, the oxidation efficiency of the fine-grained ejecta likely depends on size. The corresponding impactor diameter for a mass $M_{tangent}$ is 80 km. The diameters of transient craters under the prevailing impact conditions are estimated at $D_{tr}$ = 290 km using Eq. (A1) if $v_{impact}$ = 15 km s$^{-1}$ and $\theta_{impact}$ = 45° from the horizontal direction. For such a large impact event, $D_{tr}$ >300 km, the water-removal efficiency $\alpha$ would not reach the stoichiometric limit, because the excavated mass reaches 10–100 $M_{tangent}$. The nonoxidized ejecta deposits, however, are expected to be re-excavated by frequent small impact events and contribute to oxygen removal. The possibility of such re-excavation is discussed in detail in Section 4.1.4.

Here, I discuss the oxidation efficiency of the fine-grained ejecta for relatively small-scale impact events of impactors < 80 km. *Ghent et al.* (2010) presented the sizes of the radar-dark haloes, $r^*$, on Venus, which are produced by deposition of fine-grained ejecta beyond the continuous ejecta, as a function of the size of the host crater, $R_{crater}$, $r^*$

~ $R_{crater}^{-0.49}$ up to $R_{crater}$ = 135 km. This size relation between $r^*$ and $R_{crater}$ is fully consistent with the theoretical scaling of *Schultz* (1992a, b) for complete entrainment of the fine-grained ejecta in the impact-generated turbulent flow [*Ghent et al.*, 2010]. Thus, the fine-grained ejecta of relatively small impactors are expected to efficiently interact with and react to the hot atmosphere, and the oxidation efficiency may reach the stoichiometric limit.

*4.1.4. Turnover of the Venusian surface caused by relatively small impact events*

In the previous section, I mentioned that the nonoxidized ejecta deposited by the impacts of impactors larger than 80 km may also contribute to oxygen removal from the atmosphere through subsequent impact events. Here, I discuss the number of turnover events of the Venusian surface, $N_{turn}$, caused by impact bombardments. Figure 8 shows the probability of the number of turnovers of the Venusian surface, $N_{turn}$, exceeding 3, as a function of $q$ and $D_{max}$ for $M_{total}$ = 4 × $10^{22}$ kg. The value of $N_{turn}$ is defined as the ratio of the cumulative surface area of the transient craters produced, $\Sigma S_{tr}$, to the surface area of Venus. The probability $P$ is calculated by $P = N(N_{turm} > 3)/N_{reduced}$, where $N(N_{turm} > 3)$ and $N_{reduced}$ are respectively the number of runs for which $N_{turn}$ exceeds 3 at the end of the calculations, and the total number of runs after the data reduction discussed in Section 3. This result clearly shows that re-excavation of the nonoxidized ejecta deposited into the hot atmosphere may occur if the impactors at the terminal stage of planetary accretion have a distribution with a large $q$ and small $D_{max}$.

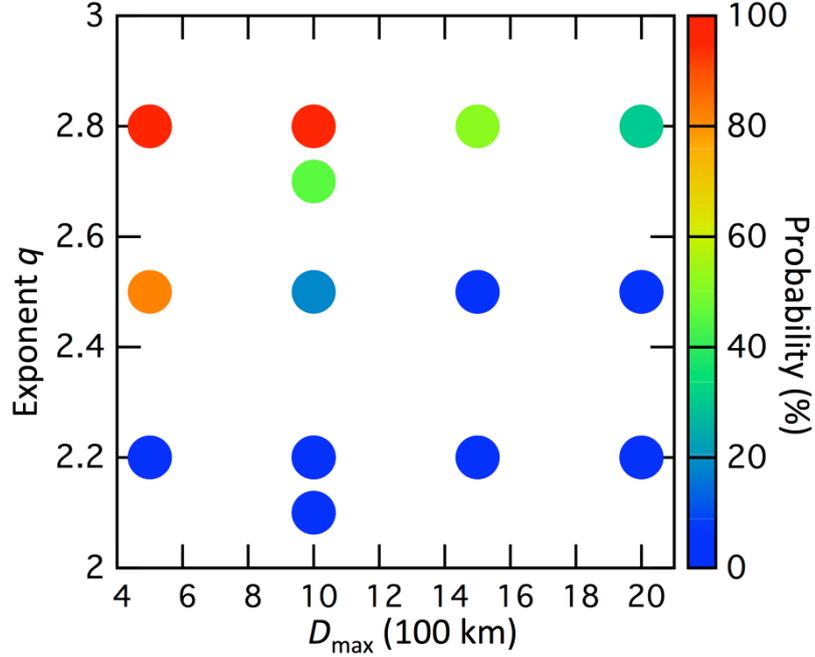

**Figure 8.** Probability of the number of turnovers of the Venusian surface, $N_{turn}$, exceeding 3, as a function of $q$ and $D_{max}$. The value of $N_{turn}$ is defined as the ratio of the cumulative surface area of the transient craters produced, $\Sigma S_{tr}$, to the surface area of Venus. The total impactor mass, $M_{total}$, was set at $4 \times 10^{22}$ kg in the calculations shown here.

### 4.2. Ferrous iron on the current Venusian surface

I assumed that the crust/mantle of Venus contains FeO of 8.6 wt % prior to the impact bombardment based on the remote-sensing data on the current Venusian surface [*Surkov et al.*, 1984, 1986] as discussed in Section 2.1. In contrast, the calculation results suggest that such surficial ferrous iron in the crust/mantle down to several tens km is oxidized to ferric iron at the terminal stage of planetary accretion, resulting in the formation of a highly-oxidized basaltic rock layer. For example, the oxidized layer 80 km in thickness is expected to be remained in the Venusian crust/mantle if the cumulative mass of fresh ejecta reaches $10^{23}$ kg. In this section, I discuss two seemingly contradictory statements. It has been widely believed that Venus may have experienced a global resurfacing event at 300-600 Myr ago based on crater counting on Venusian surface [e.g., *Phillips et al.*, 1992]. The pre-existing topographical features and geochemical signatures may have been erased by an extensive lava flow at that time. The depth of magma source is estimated to be deeper than 80 km, which is deeper than the depth of the complete excavated layer as shown in Figure 5, based on the geophysical analysis of possible plume sites on Venus [e.g., *Nimmo and McKenzie*, 1998]. Thus, the massive oxidized basalt produced by the impact bombardment is expected to be hid by the nonoxidized fresh lava flow.

**4.3. Dry interior of present-day Venus**

The fate of surface water on Venus is the main topic of this study. It is widely thought that the interior of present-day Venus is also dry, in order to explain the current volcanic and tectonic features on Venus, including its mantle viscosity, melt generation rates on possible plume sites, and the lithospheric and elastic thickness [e.g., *Nimmo* *and McKenzie*, 1998]. Unfortunately, a dry interior of Venus is not a collateral consequence of "impact-driven planetary desiccation" and exploration of this issue is beyond the scope of this study.

### 4.4. Effect of water delivery by impactors

Although the chemical composition of impactors at the terminal stage of planetary accretion has not been well constrained, impactors may deliver water to a proto-Venus. For example, the abundance of water in CI chondrites has been reported to range from 5 wt% to 20 wt% [e.g., *Kerridge*, 1985]. If all impactors were CI-like wet materials, the total amount of water delivered may reach ~1 TO. However, CI-like materials also contain a large amount of reduced species, including organic carbon, metallic iron, FeO, and FeS [e.g., *Wasson and Kallemeyn*, 1988]. Consequently, impact-delivered water might be removed efficiently by such reducing agents through fast $H_2$ escape, as discussed in Section 1.

### 4.5. The difference between Earth and Venus

I showed that the surface water up to >0.6 TO could be removed from Venus during the terminal stage of planetary accretion at $M_{total} > 2 \times 10^{22}$ kg as discussed in Section 3. The Earth must suffer similar levels of impact bombardment. In this section, I address the difference between Earth and Venus at that time. Since the semimajor axis of the Earth is larger than the Venus, a large amount of surface water has been condensed into an ocean [e.g., *Matsui and Abe*, 1986]. Thus, the photodissociation of water vapor and the accumulation of oxygen would not occur on the early Earth, suggesting that the oxidation of excavated rocks as discussed in Section 2.1 would not proceed efficiently. In other words, the form of surface water (i.e., ocean versus a steam atmosphere) is an important factor in determining the fate of surface water against to the impact bombardment. One of the important assumptions of this study, which is proto-Venus has a thick steam atmosphere, has been supported by recent atmospheric model with a modern spectral database [e.g., *Goldblatt et al.*, 2013].

**4.6. Impact driven mixing between planetary mantle and atmosphere**

The impact bombardment would be the missing driving force of the efficient mixing with the deep interior of the planetary crust/mantle and the atmosphere in a number of previous studies. The proposed process could play an important role in planetary atmospheric evolution, not only in the context of Venus but also regarding that of extrasolar planets. *Sleep and Zahnle* (2001) pointed out that chemical weathering of rocky ejecta is the main contributor in cation injection into the ocean on the early Earth at the Hadean era, resulting in ejecta-dominated climate via the fixation of atmospheric $CO_2$ into carbonates. I demonstrate that the impact-generated ejecta mass reaches up to 1 wt% of the host planet's mass for terrestrial planets in the Solar System as discussed in Section 3. The cumulative mass of the excavated rocks corresponds to >$10^4$ times of that of the current Earth's atmosphere. The removed mass of accumulated oxygen corresponds to >140 atm. Thus, heterogeneous chemical reactions between significant amounts of minerals and planetary atmospheres are expected to significantly affect the atmospheric mass and composition and would enhance the diversity of terrestrial planets, depending on the impact history during accretion of the planetary system. More detailed studies of the chemical interactions between minerals and volatiles under a wide range of pressure–temperature conditions would provide useful information in our quest to understand the atmospheric evolution of terrestrial planets.

## 5. Conclusions

I propose a new process for the removal of water vapor from proto-Venus, "impact-driven planetary desiccation". The accumulation of oxygen would occur on proto-Venus due to the photodissociation of a steam atmosphere and the subsequent hydrodynamic escape of hydrogen at the early stage of the evolution of the surface environment on Venus. Hypervelocity impacts contribute to the removal of the accumulated oxygen through oxidation of iron-bearing rocky ejecta. I have conducted the first quantitative calculations of the cumulative mass of impact-generated fine-grained ejecta on proto-Venus at the terminal stage of planetary accretion. I show that the oxidation reaction from ferrous to ferric iron proceeds in the presence of excess oxygen and a hot environment on a proto-Venus, using thermochemical calculations. In the calculations, I assumed that the fine-grained ejecta are completely oxidized to ferric iron, based on results of high-temperature oxidation experiments of iron-bearing minerals and considering the dynamics of the fine-grained ejecta in a hot and thick atmosphere on a proto-Venus. Although the estimate is an upper limit, this new process may revive the crust/mantle oxidation hypothesis to remove the surface water from the proto-Venus. Part of the Venusian crust and/or mantle down to several hundred km is excavated and disperses into the atmosphere at that time. Such efficient mixing between the planetary crust/mantle and the atmosphere may largely affect the mass and chemical composition of the atmosphere because the cumulative mass of rocky ejecta reach $10^4$ times of the atmospheric mass on the current Earth. Hypervelocity impacts during the terminal stage of planetary accretion may play important roles in the atmospheric evolution, not only in the Solar System but also in extrasolar systems.

**Acknowledgements** I am grateful to Hiroki Senshu, Takanori Sasaki, Keiko Hamano, Hiroyuki Kurokawa, Takashi Ito, Koji Wada, George L. Hashimoto, Tomohiro Usui, Masashi Ushioda, Takayuki Hirai, Norimune Miyake, and Toshihiko Kadono for helpful discussions. I thank Natsuki Hosono, Ryou Ohsawa, and Takayuki Muto for their helpful comments on the visuallization of the calculation results. I also appreciate H. Jay Melosh and two anonymous referees for their critical reviews that helped improve the manuscript greatly and Bernard Marty for helpful suggestions as an editor. I appreciate for useful discussion in the workshop on planetary impacts held at the Institute of Low Temperature Science, Hokkaido University. This research was supported by JSPS KAKENHI Grant Numbers 24244071, 25871212, and 26610184.

**Appendix A. The excavated mass and depth following each impact event**

This section describes the calculations of the mass of the impact ejecta following each hypervelocity impact. The diameter of the transient crater $D_{tr}$ is given based on the π-group scaling laws [e.g., *Abramov et al.*, 2012; *Schmidt and Housen*, 1987],

$$D_{tr} = 2R_{tr} = 0.806 \times (3.22)^{-\beta} C_D R_{proj}^{1-\beta} g^{-\beta} \left[ v_{impact} \sin(\vartheta_{impact}) \right]^{2\beta}, \quad (A1)$$

where $R_{tr}$, $\beta = 0.22$, $C_D = 0.8$, $g = 8.87$ m s$^{-2}$, $R_{proj}$ and $\theta_{impact}$ are, respectively, the radius of the transient crater, an exponent associated with π-group scaling, a constant, the gravitational acceleration on the surface of the Venus, the radius of the impactor, and the impact angle, respectively. The values of $\beta$ and $C_D$ were chosen because they have been validated as scaling constants for gravity-dominated crater formation by both experiments [*Schmidt and Housen*, 1987] and hydrocode calculations [*Elbeshausen et al.*, 2009]. The excavated volume is calculated using Maxwell's Z model, with an effective centre of flow at a constant depth (Z-EDOZ model) [*Croft*, 1980; *Stewart and Valiant*, 2006]. The formulation and good schematic representations of the excavated volume in the Z-EDOZ model can be found in the original study [*Croft*, 1980]. Therefore, only the key equations are described here. The burial depth of the effective centre of the Z-flow field physically corresponds to the depth of the isobaric core, $d_{ic}$, in the compressed region generated by the shock wave from the impact point [e.g., *Stewart and Valiant*, 2006]. The empirical definition of $d_{ic}$ as a function of impact velocity, $v_{impact}$, based on a series of hydrocode calculations [*Pierazzo et al.*, 1997] is

$$\log_{10}(d_{ic}) = -0.606 + 0.408 \times \log_{10}[v_{impact} \sin(\theta_{impact})]. \tag{A2}$$

Note that the vertical component of the impact velocity $v_{impact} \sin(\theta_{impact})$ is used to apply this equation to oblique impacts in this study. The excavation depth in the Z-EDOZ model, $H_{exc}$, is given by [*Croft*, 1980]

$$H_{exc} = d_{ic} + \frac{R_{tr}(Z-2)[Z-1]^{(1-Z)/(Z-2)}}{\left[\cos\Delta(1+\sin\Delta)^{1/(Z-2)}\right]}, \tag{A3}$$

where $Z$ and $\Delta$ are a constant to determine the curvature of the flow field and the angle between the horizontal plane of the effective centre of the cratering flow at $d_{ic}$ and the point where the streamline intersects the pre-impact target surface, respectively. The excavated volume $V_{exc}$ is described by

$$V_{exc} = \frac{\pi}{3} d_{ic} R_{tr}^2 + \frac{2\pi}{3} R_{tr}^3 \frac{(Z-2)}{(Z+1)} \frac{1+\sin\Delta}{\cos^3\Delta}. \quad (A4)$$

I employed $Z = 3$ because this choice appears to be a good approximation to actual cratering flows in the gravity-dominated regime [e.g., *Melosh*, 1989]. In this case, Eqs. (A3) and (A4) can be expressed in much simpler forms,

$$H_{exc} = d_{ic} + \frac{R_{tr}}{[4\cos\Delta(1+\sin\Delta)]} \quad \text{and} \quad (A5)$$

$$V_{exc} = \frac{\pi}{3} d_{ic} R_{tr}^2 + \frac{\pi}{6} R_{tr}^3 \frac{1+\sin\Delta}{\cos^3\Delta}. \quad (A6)$$

Eqs (A5) and (A6) were used in the cratering calculations presented in this study. **(428 words)**

early Earth, *Journal of Geophysical Research*, **103**, 28,529–28,544.

Sleep, N. H. and Zahnle, K. 2001, Carbon dioxide cycling and implications for climate on ancient Earth, *Journla of Geophysical Research*, **106**, 1373-1399.

Smit, J. 1999, The global stratigraphy of the Cretaceous–Tertiary boundary impact ejecta, *Annual Review of Earth and Planetary Sciences*, **27**, 75–113.

Stewart, S. T. and Valiant, G. J. 2006, Martian subsurface properties and crater formation processes inferred from fresh impact crater geometries, *Meteoritics and Planetary Science*, **41,** 1509–1537.

Surkov, Y. A., Barsukov, V. L., Moskalyeva, L. P., Kharyukova, V. P. and Kemurdzhian, A. L. 1984, New data on the composition, structure, and properties of Venus rock obtained by Venera 13 and Venera 14, *Journal of Geophysical Research*, **89,** B393.

Surkov, Y. A., Moskalyova, L. P., Kharyukova, V. P., Dudin, A. D., Smirnov, G. G., and Zaitseva, S. Y. 1986, Venus rock composition at the Vega 2 Landing Site, *Journal of Geophysical Research*, **91,** E215.

Tanaka, H., Inaba, S., and Nakazawa, K. 1996, Steady-state size distribution for the self-similar collision cascade, *Icarus*, **123**, 450–455.

Terada, N., Kulikov, Y. N., Lammer, H., Lichtenegger, H. I. M., Tanaka, T., Shinagawa, H., and Zhang, T. 2009, Atmosphere and Water Loss from Early Mars Under Extreme Solar Wind and Extreme Ultraviolet Conditions, *Astrobiology*, **9**, 55–70.